\documentclass[12pt,letterpaper]{article}
\usepackage[affil-it]{authblk}
\usepackage{amsmath, amsfonts, amsthm, amssymb}
\usepackage[usenames,dvipsnames]{color}
\usepackage{subcaption}
\usepackage{comment}
\usepackage{ifpdf}
\usepackage{setspace}
\usepackage{bbold}
\usepackage[utf8]{inputenc}
\usepackage[english]{babel}
\usepackage{graphicx}  
\usepackage{breqn}
\usepackage{mathrsfs}
\usepackage{cite}
\usepackage[left=.8in,right=.8in,top=1in,bottom=1in]{geometry}  
\newcommand*\colvec[3][]{
    \begin{pmatrix}\ifx\relax#1\relax\else#1\\\fi#2\\#3\end{pmatrix}
}
\newcommand{\overbar}[1]{\mkern 1.5mu\overline{\mkern-1.5mu#1\mkern-1.5mu}\mkern 1.5mu}
\newcommand{\pder}[2]{\frac{\partial#1}{\partial#2}}
\begin{document}
\title{\textbf{Parity Horizons in Shape Dynamics}}   
\author{Gabriel Herczeg\footnote{email: Herczeg@ms.physics.ucdavis.edu}}
\affil{Department of Physics, \\ University of California, Davis \\ Davis California, 95616}

\makeatletter
\renewcommand*{\p@section}{\S\,}
\renewcommand*{\p@subsection}{\S\,}
\makeatother

\makeatletter
\renewcommand*{\p@figure}{Fig.\,}
\makeatother

\date{\today}      
\maketitle

\begin{abstract} 
\noindent I introduce the notion of a parity horizon, and show that many simple solutions of shape dynamics possess them. I show that the event horizons of the known asymptotically flat black hole solutions of shape dynamics are parity horizons and that this notion of parity implies that these horizons possess a notion of CPT invariance that can in some cases be extended to the solution as a whole. I present three new solutions of shape dynamics with parity horizons and find that not only do event horizons become parity horizons in shape dynamics, but observer-dependent horizons and Cauchy horizons do as well. The fact that Cauchy horizons become (singular) parity horizons suggests a general chronology protection mechanism in shape dynamics that prevents the formation of closed timelike curves.
\end{abstract}

\pagebreak

\section{Introduction}
\subsection{Background}\label{Background}
Shape dynamics is a classical theory of gravity that shares many features in common with the canonical formulation of general relativity due to Arnowitt, Deser and Misner \cite{ADM}, and which makes identical predictions about the outcomes of the experiments which have confirmed general relativity to date\footnote{Readers seeking additional background on shape dynamics should refer to \cite{GGK, H-Thesis, tutorial, FAQ, York, gravDOF, TimMatter}.}. 
Originally \cite{GGK}, shape dynamics was viewed as a reformulation of canonical general relativity that traded the (on-shell) refoliation invariance generated by the quadratic Hamiltonian constraint for spatial Weyl invariance generated by a new, linear Weyl constraint. The theories were quickly seen to agree for a broad class of known solutions to general relativity---particularly those admitting globally defined foliations by spatial hypersurfaces of constant mean extrinsic curvature\footnote{For compact spatial manifolds, shape dynamics agrees with canonical general relativity when spacetime is foliated by hypersurface surfaces of constant mean extrinsic curvature. For asymptotically flat spatial manifolds, such as those discussed in this paper, shape dynamics agrees with general relativity when spacetime is foliated by maximal slices---meaning that the trace of the momentum conjugate to spatial metric vanishes.}. However, questions remained about what the corresponding solutions of shape dynamics would look like for spacetimes that admitted such foliations only locally, or whether it was even possible for shape dynamics to describe such systems.

A partial answer to this question was provided by \cite{Birkhoff} and \cite{Kerr}, in which spherically symmetric and rotating asymptotically flat black hole solutions of shape dynamics were presented. These novel solutions were shown to agree with their general relativistic counterparts outside their event horizons, but behave differently in their interior regions. Both solutions possess an inversion symmetry about their event horizons, which implies that the interior regions should be interpreted as a time-reversed copy of the exterior regions, endowing the solutions with the character of a \emph{wormhole}. I will show in the following sections that these features are not peculiar to event horizons, but arise naturally in solutions of shape dynamics with other types of horizons as well. I analyze the properties of these horizons and show that all of them are parity horizons, which are defined in \ref{SDBH}. 

One way to understand the disagreement between shape dynamics and general relativity when solutions develop horizons is to note that in the canonical formalism, many types of horizons arise when the lapse function either vanishes or diverges. Since the determinant of the spacetime metric can be written $\sqrt{-g} = N\sqrt{q}$ where $q$ is the determinant of the spatial metric and $N$ is the lapse function, it is clear that on a horizon where the lapse vanishes or diverges, the determinants of the spatial and spacetime metrics cannot simultaneously be finite and non-zero---one must choose between a smooth spacetime and a smooth conformal spatial geometry. 

With the benefit of hindsight, it is not particularly surprising that shape dynamics and general relativity disagree when solutions of either develop horizons, however it is still interesting to consider the particular ways in which the theories disagree in the presence of such surfaces.

The remainder of the paper is organized as follows. In the following section, I will review the known asymptotically flat black hole solutions for shape dynamics  and discuss their properties. In \ref{charge}, I introduce a spherically symmetric, charged black hole solution for shape dynamics analogous to the Reissner-N\"{o}rdstrom black hole of general relativity, and discuss the properties of the solution. In \ref{rindler}, I discuss the Rindler chart over Minkowski spacetime and introduce a new solution of shape dynamics which shares many of the same features. I will discuss how this new solution differs from its general relativistic counterpart and emphasize the crucial role played by the boundary conditions in shape dynamics. In \ref{bonner}, I briefly review closed timelike curves and discuss a family of solutions to Einstein's equations known as the Van Stockum-Bonner spacetimes, which except for the lowest order case, are stationary, asymptotically flat and possess a compact\footnote{Here, ``compact" means that the Cauchy horizon is compact when viewed as a (degenerate) two-surface embedded in any of the preferred spatial slices.}  Cauchy horizon within which closed timelike curves pass through every point. I will then present a novel solution of shape dynamics which agrees with the next-to-lowest order Bonner spacetime outside the horizon, but which does not develop closed timelike curves within---the Cauchy horizon is replaced by a parity horizon in the shape dynamics solution. I argue that this is evidence of a general chronology protection mechanism in shape dynamics that is significantly more parsimonious than many of the arguments that have been made for chronology protection in general relativity. 

\subsection{Shape Dynamics}
 Before continuing to the main results, it is worthwhile to review the construction of shape dynamics for asymptotically flat spatial manifolds presented in \cite{H-thesis} and elaborated in \cite{poincare}. One begins with the standard canonical formulation of general relativity. The canonical form of the Einstein-Hilbert action is given by

\begin{equation}\label{ADM action}
\mathcal{I}_{\mbox{\tiny ADM}} = \int dt\bigg[ \frac{1}{16\pi}\int_{\Sigma} d^3x\sqrt{q} \left(\dot{q}_{ij}\pi^{ij} -
  N\mathcal{S}(x) - \xi^i\mathcal{H}_i(x)\right)  -
\frac{1}{8\pi}\int_{\partial\Sigma} d^2x\sqrt{\sigma}(NK+r^iN_{,i}-r_i\xi^j\pi^i_j) \bigg] 
\end{equation}

\noindent where $\pi^{ij} = \frac{\partial\mathcal{L}}{\partial \dot{q}_{ij}}$ is the momentum  canonically conjugate to $q_{ij}$, $\sigma_{ab}$ is the metric induced on $\partial\Sigma$ by $q_{ij}$, $r^i$ is the outward pointing normal vector of  $\partial\Sigma$, $K$ is the trace of the extrinsic curvature of $\partial\Sigma$ embedded in $\Sigma$, and $N$ and $\xi^i$ are the lapse function and shift vector. The quantities $\mathcal{S}(x)$ and $\mathcal{H}_a(x)$ are the scalar constraint (or ``Hamiltonian constraint") and the diffeomorphism constraint (or ``momentum constraint"), defined by

\begin{eqnarray}\label{ADM const.}
\mathcal{S}(x) &=& \frac{G_{ijkl}\pi^{ij}\pi^{kl}}{\sqrt{q}} - R\sqrt{q} \\
\mathcal{H}_i(x) &=& -2\pi^j_{i;j}
\end{eqnarray}

The scalar constraint\footnote{The quantity $G_{ijkl}$ appearing in the scalar constraint is the DeWitt supermetric defined by: \\ $G_{ijkl}:= \frac{1}{2}(q_{ik}q_{jl} + q_{il}q_{jk})-q_{ij}q_{kl}.$} $\mathcal{S}(x)$ generates refoliations of spacetime provided the equations of motion are satisfied, and the diffeomorphism constraint $\mathcal{H}_i(x)$ generates foliation-preserving spatial diffeomorphisms.

The next step in the construction of shape dynamics is to extend the phase space of canonical general relativity by a scalar field $\phi$ and its canonically conjugate momentum $\pi_{\phi}$, where $e^{4\phi}$ plays the role of a ``conformal factor" satisfying the fall-off condition $e^{4\phi} \sim 1 + \mathcal{O}(r^{-1})$. Together with the new first class constraint $\pi_{\phi} \approx 0$, the addition of these new canonical variables trivially embeds canonical general relativity into the extended phase space $(q_{ij}, \pi^{ij},\phi, \pi_{\phi})$. 

To obtain a non-trivial embedding, one then performs the canonical transformation 

\begin{eqnarray}
T_{\phi}q_{ij}(x) &:=& e^{4\phi(x)}q_{ij}(x) \nonumber \\
T_{\phi}\pi^{ij}(x) &:=& e^{-4\phi(x)}\pi^{ij}(x) \nonumber \\
T_{\phi}\phi(x) &:=& \phi(x) \nonumber \\
T_{\phi}\pi_{\phi}(x) &:=& \pi_{\phi}(x)-4q_{ij}(x)\pi^{ij}(x) \nonumber
\end{eqnarray}

\noindent The Hamiltonian for the extended phase space can be written in terms of the canonically transformed first class constraints:

\begin{equation}\label{H-linking}
\mathcal{H}_{\mbox{\tiny link}} = \int_{\Sigma}d^3x\left[N(x)T_{\phi}\mathcal{S}(x) + \xi^i(x)T_{\phi}\mathcal{H}_i(x) + \rho(x)T_{\phi}\pi_{\phi}(x)\right] + T_{\phi}B(N,\xi)
\end{equation}

\noindent where $N(x)$ is the lapse function, $\xi^a(x)$ is the shift vector and $\rho(x)$ is a Lagrange multiplier for the Weyl constraint $T_{\phi}\pi_{\phi}$. $B(N,\xi)$ is the Gibbons-Hawking-York boundary term (the surface integral in \eqref{ADM action}), and $T_{\phi}B(N,\xi)$ plays an important role in defining the globally conserved charges in shape dynamics \cite{poincare}. The system described by this Hamiltonian is known as the \emph{linking theory}, as it provides a link between canonical general relativity and shape dynamics \cite{Linking}. In order to obtain shape dynamics from the linking theory, one imposes the gauge fixing condition $\pi_{\phi}(x) \equiv 0$. The only constraint whose Poisson bracket with the gauge-fixing condition $\pi_{\phi}(x) \equiv 0$ is weakly non-vanishing is $T_{\phi}\mathcal{S}(N)$:

\begin{equation}\label{weak bracket}
\{T_{\phi}\mathcal{S}(N),\pi_{\phi}(x)\} = 4T_{\phi}\{\mathcal{S}(N),\pi(x)\}
\end{equation}

\noindent where $\mathcal{S}(N) = \int_{\Sigma}d^3xN(x)S(x)$ is the scalar constraint smeared with the lapse function $N(x)$, and $\pi(x) = q_{ij}(x)\pi^{ij}(x)$. Equation \eqref{weak bracket} implies the ``lapse-fixing equation":

\begin{equation}\label{LFE1}
T_{\phi}\{\mathcal{S}(N), \pi(x)\} \approx
e^{-4\phi}(\nabla^2 N + 2q^{ij}\phi_{,i}N_{,j}) -
e^{-12\phi}NG_{ijkl}\frac{\pi^{ij}\pi^{kl}}{|q|} \approx 0
\end{equation}

\noindent where $R$ is the scalar curvature of $\Sigma$, and $\approx$ denotes ``weak equality"---i.e. equality up to an additive term that vanishes when the constraints are satisfied.  The solution $N_0(x)$ of the lapse-fixing equation\footnote{For a physical interpretation of the solution of \eqref{LFE1} as an effective ``experienced lapse" for weak matter field fluctuations see \cite{TimMatter}. For a related discussion of how spacetime emerges from coupling shape dynamics to matter see \cite{SD-Matter}.} \eqref{LFE1} is unique up to a choice of boundary conditions on the lapse. This will be discussed in further detail in \ref{rindler}.

After imposing the gauge fixing condition $\pi_{\phi}(x) \equiv 0$ and working out the consistency conditions in the algebra of constraints, we are left with the first class constraints $T_{\phi}\mathcal{S}(N_0)$, $-4\pi(x)$ and $T_{\phi}\mathcal{H}_i(x) = \mathcal{H}_i(x)$ and the second class constraints $T_{\phi}\mathcal{S}(x)-T_{\phi}\mathcal{S}(N_0)\sqrt{q}(x)$ and $\pi_{\phi}(x)$. Of the remaining first class constraints, $T_{\phi}\mathcal{S}(N_0)$ generates time reparametrizations, $\mathcal{H}_i(x)$ generates diffeomorphisms acting on the conformal spatial metric and it conjugate momentum, and $-4\pi(x)$ generates spatial Weyl transformations. To summarize, the total Hamiltonian for shape dynamics is given by 

\begin{equation}\label{shapeHam}
\mathcal{H}_{\mbox{\tiny SD}} = T_{\phi}S(N_0) + \mathcal{H}_i(\xi^i) -4\pi(\rho)
\end{equation}

\noindent which leads to the first-order equations of motion:

\begin{eqnarray}\label{EOM}
\dot{q}_{ij} &=& 4\rho q_{ij} + 2e^{-6\phi}\frac{N_0}{\sqrt{q}}\pi_{ij} + \mathcal{L}_{\xi}q_{ij}  \\
\dot{\pi}^{ij} &=& N_0 e^{2\phi}\sqrt{q}(R^{ij} - 2 \phi^{;ij} + 4 \phi^{;i}\phi^{;j} - \frac{1}{2}Rq^{ij} + 2\nabla^2\phi q^{ij}) \nonumber  \\
 & & -e^{2\phi}\sqrt{q}(N_0^{;ij} - 4\phi^{(,i}N_0^{,j)} - \nabla^2 N_0q^{ij}) + \mathcal{L}_{\xi}\pi^{ij} - 4\rho\pi^{ij}  \\
 & & -\frac{N_0}{\sqrt{q}}e^{-6\phi}(2\pi^{ik}\pi^{j}_{k} - \pi^{kl}\pi_{kl}q^{ij}). \nonumber \\ \nonumber
\end{eqnarray}

While the asymptotically flat formulation of shape dynamics described above is very useful for studying isolated systems, it should be thought of as an approximation to the more fundamental, spatially compact formulation of the theory. In that spirit, the solutions studied here should be thought of as approximately describing nearly empty regions of a much larger, spatially compact universe. The primary drawback of the asymptotically flat framework is that the solutions depend on a choice of boundary conditions, which spoils some of the relationalism  that the full, spatially compact theory boasts \cite{SDanIntro}. Moreover, it is not yet clear what physical role the degeneracy introduced by the freedom to choose boundary conditions plays in the theory. The reader should particularly keep this point in mind when reading \ref{rindler}.
\section{Shape Dynamic Black Holes}\label{SDBH}
Assuming asymptotically flat and Lorentz-invariant boundary conditions\footnote{For a more general analysis of spherically symmetric, asymptotically \emph{spatially} flat solutions of shape dynamics that are not necessarily asymptotically Lorentz-invariant see \cite{FateOfBirkhoff}.}, shape dynamics admits a unique spherically symmetric solution that can be described globally by the Schwarzschild line element in isotropic coordinates:

\begin{equation}\label{isotropic}
ds^2= -\left(\frac{1-\frac{m}{2r}}{1+\frac{m}{2r}}\right)^2dt^2+ \left(1+\frac{m}{2r}\right)^4\left(dr^2+r^2(d\theta^2+\sin^2\theta d\phi^2)\right).
\end{equation}

\noindent This line element is a solution of the Einstein field equations \emph{only} for $r \neq m/2$, where $m/2$ is the location of the event horizon in isotropic coordinates, whereas this solution is valid \emph{everywhere} from the point of view of shape dynamics. The line element is invariant under the transformation  $r\to m^2/(4r)$. It can be seen from this inversion symmetry is that the ``interior" of this solution is a time-reversed copy of the exterior, which makes manifest the wormhole character of the solution and that the solution possesses no central curvature singularity. In fact, from the point of view of shape dynamics, this solution is completely free of physical singularities since there are no singularities in the spatial conformal geometry. This can be seen by noting that the Cotton tensor\footnote{The Cotton tensor is defined by $\mathscr{C}_{ijk} := \nabla_k\left(R_{ij} - \frac{1}{4}Rg_{ij}\right) - \nabla_j\left(R_{ik} - \frac{1}{4}Rq_{ik}\right)$. The Cotton tensor possesses all of the local information on the conformal structure of a three-dimensional Riemannian manifold just as the Weyl tensor contains all of the local information on the conformal structure of higher-dimensional manifolds \cite{Gourgoulhon}.} vanishes identically, which follows from the fact that the spatial metric is conformally flat.

\begin{figure}[h]
\begin{subfigure}{.5\textwidth}
  \centering
  \includegraphics[width=.8\linewidth]{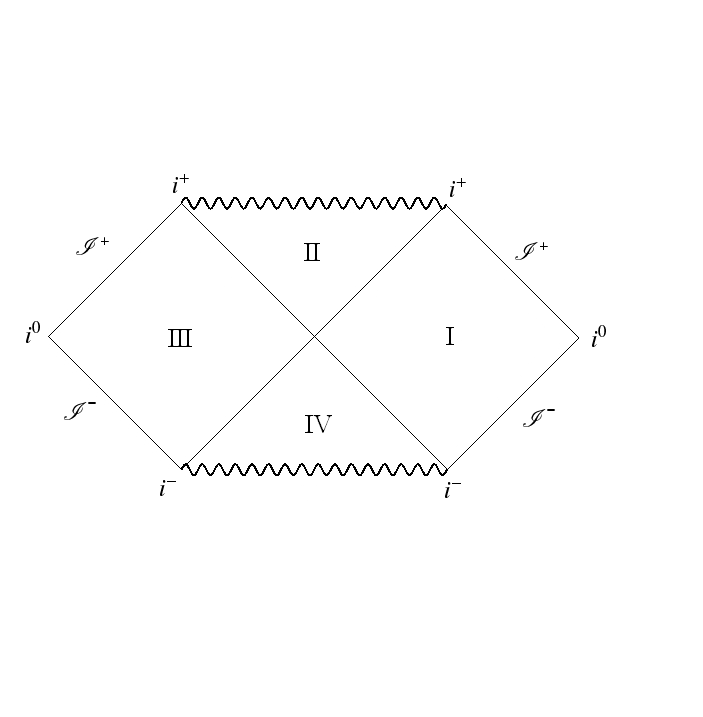}
  \caption{\,}
  \label{penrose-GR}
\end{subfigure}%
\begin{subfigure}{.5\textwidth}
  \centering
  \includegraphics[width=.8\linewidth]{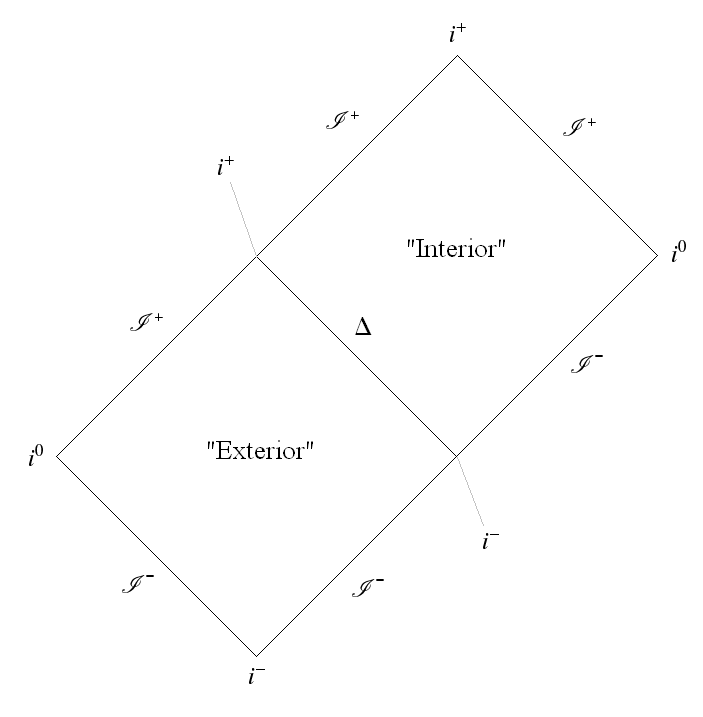}
  \caption{\,}
  \label{penrose-SD}
\end{subfigure}
\caption{\footnotesize On the left, (a) shows the conformal diagram of the maximally extended Kruskal spacetime. On the right, (b) shows the conformal diagram for a spherically symmetric, shape dynamic black hole. In each, $i^{\pm}$ are future and past null infinity, $i^0$ is spatial infinity, $\mathscr{I}^{\pm}$ are future and past null infinity and $\Delta$ is the event horizon.}
\label{penrose}
\end{figure}

This solution can be understood in relation to the maximally extended Kruskal spacetime by considering the conformal diagrams displayed in \ref{penrose}. The conformal diagram of the Kruskal spacetime (Fig. \ref{penrose-GR}) contains four regions. Regions I and III are causally disconnected and each possesses it's own spatial infinity. Region II containing a future space-like singularity and region IV contains a past space-like singularity. On the other hand, the conformal diagram of the spherically symmetric shape dynamic black hole (Fig. \ref{penrose-SD}) contains only the two regions labeled (somewhat arbitrarily) ``interior" and ``exterior." Much like regions I and III of the conformal diagram of the Kruskal spacetime, the interior and exterior regions each possess their own spatial infinity. However, unlike the Kruskal spacetime, the interior and exterior regions meet along the event horizon $\Delta$, and are therefore not causally disconnected. Observers can pass in one direction across the horizon from the exterior region to the interior. The regions containing the past and future singularities are absent in the shape dynamics case, reflecting the global regularity of the solution.

The inversion symmetry of this black hole solution is known and has been discussed in \cite{Birkhoff} and \cite {Kerr}. This property can be used to motivate the definition of a parity horizon which will be referenced throughout the paper. 
\\
\begin{quote}
\noindent Definition: Let $\{N, \xi, q_{ij}, \pi^{ij}\}$ be a solution of the equations of motion of shape dynamics in the gauge in which $\rho(x) = 0$, $e^{4\phi(x)} = 1$ and let $\dot{q}_{ij} = 0$. If $q_{ij}$ admits a two-surface $\mathcal{S}_0$ on which the lapse function $N(x)$ either vanishes or diverges and if there exists a spatial isometry $\mathcal{P}$ such that:
\begin{enumerate}
\item
 $\mathcal{P}^*q = q$ 
\item
 $\mathcal{P}\circ\mathcal{P} = \mathbb{1}$ 
\item
 $\mathcal{P}^*N = -N$.
\end{enumerate}
Then  $\mathcal{P}$ is called a parity and $\mathcal{S}_0$ is called a parity horizon. \\
\end{quote}

It is possible that this definition might be improved upon by restating it in a gauge invariant manner and/or extending it to include solutions with $\dot{q}_{ij} \neq 0$. However, for our purposes the above definition is adequate. Clearly, the event horizon of the spherically symmetric black hole solution discussed above is a parity horizon. In the remainder of this paper, I will consider various examples of known and novel solutions of shape dynamics which possess parity horizons, and discuss how the presence of this novel feature leads to physical differences between solutions of shape dynamics and solutions of general relativity that agree outside of horizons but disagree at and within them.

Let us now consider another shape dynamic black hole solution whose event horizon is a parity horizon. It was shown in \cite{Kerr} that shape dynamics admits a rotating black hole solution that also possesses an inversion symmetry about the horizon manifesting its wormhole character, and it was argued that this solution is also free of physical singularities. The reconstructed line element associated with this solution can be written:

\begin{equation}\label{Kerr PS} 
ds^2 = -\lambda^{-1}\left(dt - \omega^{-1}d\phi\right)^2 + \lambda \left[m^2e^{2\gamma}(d\mu^2 + d\theta^2) + s^2d\phi^2\right]
\end{equation}
where 
\begin{eqnarray}\label{Kerr functions PS} 
s \hspace{5pt} &=& mp\sinh\mu\sin\theta \nonumber \\
e^{2\gamma} &=& p^2\cosh^2\mu  + q^2\cos^2\theta - 1 \nonumber\\
\omega^{-1} \hspace{5pt} &=& e^{-2\gamma}\left[2mq\sin^2\theta(p\cosh\mu + 1)\right]  \\
\lambda \hspace{5pt} &=& e^{-2\gamma}\left[(p\cosh\mu + 1)^2 + q^2\cos^2\theta \right]. \nonumber \\ 
p^2 + q^2 &=& 1 \nonumber
\end{eqnarray} 

The solution becomes spherically symmetric when $p \to 1$, $q \to 0$. The location of the horizon in these coordinates is $\mu = 0$ and the transformation $\mu \to -\mu$ leaves the line element invariant, so this solution also possesses a parity about the horizon. In addition to avoiding the ring curvature singularity that is present in the interior of the Kerr spacetime, this solution also avoids the closed timelike curves present in that region. Since shape dynamics also has a preferred time variable, it is natural to wonder if it is a general feature of solutions of shape dynamics that closed timelike curves do not occur. This may indeed be the case, and in \ref{bonner} I will show that this is true for another solution of shape dynamics whose general relativistic counterpart contains closed timelike curves but possesses no event horizon. I will argue based on the generic features of this solution that chronology protection may be a general feature of the theory.

In the following section, I introduce a charged black hole solution for shape dynamics analogous to the Reissner-N\"{o}rdstrom black hole in general relativity. It is shown that the event horizon is again a parity horizon and that the parity about the horizon leads to a fully CPT invariant solution.

\section{A Reissner-N\"{o}rdstrom-Like Shape Dynamic Black Hole}\label{charge}
To address the question of whether charged shape dynamic black holes possess some notion of CPT invariance, one must first couple the linking theory to the electromagnetic field and perform a phase space reduction to obtain shape dynamics coupled to electromagnetism \cite{H-thesis}. The Hamiltonian density for electromagnetism is given by 

\begin{equation}\label{emHam}
\mathcal{H}_{\mbox{\tiny EM}} = 2\left(A_{[i,j]}A_{[k,l]}q^{ik}q^{jl}\sqrt{q} + \frac{\tilde{E}^i\tilde{E}^jq_{ij}}{\sqrt{q}}\right)
\end{equation}

\noindent where $A_i$ is the vector potential and $\tilde{E}^i$ is its canonically conjugate momentum. The physical electric field $E^i$ is related to the vector density $\tilde{E}^i$ by $E^i = \frac{\tilde{E}^i}{\sqrt{q}}$. Coupling electromagnetism to general relativity yields the system of constraints:

\begin{eqnarray}\label{coupled constraints}
\mathcal{S}(x) &=& \frac{\pi^{ij}\pi_{ij} - \pi^2}{\sqrt{q}} - \sqrt{q}R + \mathcal{H}_{\mbox{\tiny EM}} \\
\mathcal{H}_a(\xi^a) &=& \int d^3x\left(q_{ij}\mathcal{L}_{\xi}\pi^{ij} + A_{i}\mathcal{L}_{\xi}\tilde{E}^i\right) \\
G(x) &=& \nabla_i\tilde{E}^i 
\end{eqnarray}

The next step is to extend the phase space and perform the canonical transformation of the constraints. Before proceeding, one must determine how $A_i$ and $E^i$ transform under the canonical transformation $T_{\phi}$. Here it is assumed as in \cite{H-thesis}, that $A_i$ and $E^i$ transform trivially under $T_{\phi}$ so that one can retain a system of first class constraints with well-defined global charges.  With this additional assumption, the transformed coupled scalar constraint becomes

\begin{equation}\label{coupled LY-1}
T_{\phi}\mathcal{S}(x) = \sqrt{q}\Omega\left(8\nabla^2\Omega - R\right) + \frac{\pi_{ij}\pi^{ij}-\pi^2}{\sqrt{q}}\Omega^{-7} + \Omega^{-2}\mathcal{H}_{\mbox{\tiny EM}} \approx 0.
\end{equation}

\noindent where $\Omega=e^{\phi}$. Imposing the gauge-fixing condition $\pi_{\phi} \equiv 0$ one obtains the coupled second class constraint 

\begin{equation}\label{coupled LY}
T_{\phi}\mathcal{S}(x) \approx \sqrt{q}\Omega\left(8\nabla^2\Omega - R\right) + \frac{\pi_{ij}\pi^{ij}}{\sqrt{q}}\Omega^{-7} + \Omega^{-2}\mathcal{H}_{\mbox{\tiny EM}} \approx 0.
\end{equation}

\noindent where $\pi(x)$ is now once again viewed as a first class constraint generating spatial Weyl transformations. As in the uncoupled case, one can also derive a consistency condition that fixes the lapse function:

\begin{equation}\label{coupled LFE}
T_{\phi}\{\mathcal{S}(N), \pi(x)\} \approx
e^{-4\phi}(\nabla^2 N + 2q^{ij}\phi_{,i}N_{,j}) -
N\left(e^{-12\phi}G_{ijkl}\frac{\pi^{ij}\pi^{kl}}{|q|} + e^{-8\phi}\frac{\mathcal{H}_{\mbox{\tiny EM}}}{\sqrt{q}}\right) \approx 0.
\end{equation}

In order to find a simple solution to the coupled constraints, we consider spherically symmetric conformal initial data which trivially satisfy the coupled first class constraints if we choose $\rho = 0 = \xi^i$:

\begin{equation}\label{initial data}
\bar{q}_{ij} = \eta_{ij}, \hspace{.5cm} \bar{\pi}^{ij} = 0, \hspace{.5cm}\bar{A_i} = 0, \hspace{.5cm} \bar{E}^i = \frac{Q}{r^2}\hspace{.05cm}\delta^i_r
\end{equation}

\noindent where $\eta_{ij}$ is the flat spatial metric written in spherical coordinates, and the barred quantities are arbitrarily rescaled according to their conformal weight by a conformal factor $\Omega$ which is to be determined by solving \eqref{coupled LY}. Note that the chosen initial data is static and spherically symmetric, so the equations of motion are trivial, and that the spherically symmetric electric field $\bar{E}^i$ is just the coulomb electric field in the flat background defined by $\bar{q}_{ij} = \eta_{ij}$.

Since the conformal initial data is written in terms of the flat spatial metric $\eta_{ij}$, the scalar curvature $R$ vanishes, and \eqref{coupled LY} becomes simply

\begin{equation}\label{LY temp1}
8\Omega^3\nabla^2\Omega + \frac{\mathcal{H}_{\mbox{\tiny EM}}}{\sqrt{q}} = 0.
\end{equation}

\noindent Putting \eqref{initial data} into \eqref{emHam} gives an explicit expression for $\mathcal{H}_{\mbox{\tiny EM}}$ which can be substituted into  \eqref{LY temp1} to give

\begin{equation}\label{LY temp2}
8\Omega^3\left(\Omega'' + \frac{2}{r}\Omega'\right) + \frac{2Q^2}{r^4} = 0
\end{equation}

\noindent where I have also used spherical symmetry to assume that $\Omega$ depends only on $r$ and primes denote differentiation with respect to $r$. Equation \eqref{LY temp2} is difficult to solve in its present form, but it can be simplified by making the substitution $\Omega^2 = \psi$, which yields

\begin{equation}\label{LY psi}
-2\left(\psi'\right)^2 + 4\psi\psi'' + \frac{8}{r}\psi\psi' + \frac{2Q^2}{r^4}=0.
\end{equation}

\noindent Now equation \eqref{LY psi} can be solved by making the Laurent series ansatz:

\begin{equation}\label{laurent}
\psi = \sum\limits_{n=0}^{\infty}c_nr^{-n}.
\end{equation}

\noindent The derivatives of $\psi$ can be easily calculated from \eqref{laurent}:

\begin{eqnarray}\label{laurent prime}
\psi' &=& -\sum\limits_{n=0}^{\infty}nc_nr^{-(n+1)} \nonumber \\
\psi'' &=& \sum\limits_{n=0}^{\infty}n(n+1)c_nr^{-(n+2)}.
\end{eqnarray}

\noindent Inserting \eqref{laurent} and \eqref{laurent prime} back into \eqref{LY psi} yields the infinite double sum:

\begin{equation}\label{double sum}
\sum\limits_{m=0}^{\infty}\sum\limits_{n=0}^{\infty}c_m c_n \left[-2mn + 4n(n-1)\right] r^{-(2 + m + n)} = -\frac{2Q}{r^4}.
\end{equation}

\noindent In order for \eqref{double sum} to be satisfied to all orders, all terms for which $m+n \neq 2$ on the left-hand side must vanish. This implies that $c_i = 0$ for $i>2$, which means that the series terminates. Collecting the terms proportional to $r^{-4}$ on the left hand side gives:

\begin{equation}\label{constants}
Q^2 + 4c_0 c_2 - c_1^2 = 0.
\end{equation}

\noindent Imposing the generic boundary conditions\footnote{The choice to rename $c_1 = m$ can be justified a fortiori by noting that the ADM mass of the resulting solution is equal to m.} $c_0 = 1$, $c_1 = m$, equation \eqref{constants} can be solved for $c_2$ in terms of the mass $m$ and electric charge $Q$:

\begin{equation}\label{c2}
c_2 = \frac{m^2-Q^2}{4}.
\end{equation}

\noindent Inserting equation  \eqref{c2} back into equation \eqref{laurent} and recalling that $c_n = 0$ for $n>2$:

\begin{equation}\label{Omega final}
\psi = 1 + \frac{m}{r} + \frac{m^2-Q^2}{4r^2} \hspace{.45cm}
\implies \hspace{.45cm}
\Omega =  \left(1 + \frac{m}{r} + \frac{m^2-Q^2}{4r^2}\right)^{1/2}.
\end{equation}

Putting equations \eqref{Omega final} and \eqref{initial data} into \eqref{coupled LFE}, one obtains the homogenous, linear, second order ordinary differential equation:

\begin{equation}\label{coupled LFE 2}
\Omega^4\left(N'' + 2\left(\frac{1}{r} + \frac{\Omega'}{\Omega\hspace{.07cm}}\right)N'\right) - \frac{Q^2}{r^4}N = 0
\end{equation}

\noindent which given asymptotically flat boundary conditions, has the unique solution:

\begin{equation}
N = \frac{1 - \frac{m^2 - Q^2}{4r^2}}{1 + \frac{m}{r} + \frac{m^2-Q^2}{4r^2}}.
\end{equation}

Rescaling the conformal initial data by the appropriate powers of $\Omega$ according to their conformal weights gives the spatial metric and physical electric field:

\begin{eqnarray}\label{rescaled initial data}
q_{ij} &=& \Omega^4\bar{q}_{ij} =  \left(1 + \frac{m}{r} + \frac{m^2-Q^2}{4r^2}\right)^2\eta_{ij} \\
E^i &=& \Omega^{-6}\bar{E}^i =  \left(1 + \frac{m}{r} + \frac{m^2-Q^2}{4r^2}\right)^{-3}\frac{Q}{r^2}\delta^i_r
\end{eqnarray}

The reconstructed line element associated with this solution of shape dynamics is given by

\begin{equation}\label{Nord element}
ds^2 = -N^2 dt^2 + q_{ij}dx^i dx^j = \left(\frac{1 - \frac{m^2 - Q^2}{4r^2}}{1 + \frac{m}{r} + \frac{m^2-Q^2}{4r^2}}\right)^2 dt^2 + \left(1 + \frac{m}{r} + \frac{m^2-Q^2}{4r^2}\right)^{2}\left(dr^2 + r^2 dS_{\mbox{\tiny 2}}^2\right)
\end{equation}

\noindent where $dS_{\mbox{\tiny 2}}^2 = d\theta^2 + \sin^2\theta d\phi^2$ is the metric on the unit two-sphere. Equation \eqref{Nord element} is just the line element of the Reissner-N\"{o}rdstrom black hole written in isotropic coordinates, which is well known in the context of general relativity, and has been derived by similar methods in for example \cite{Alcubierre}. The physical difference between this shape dynamics solution and the Reissner-N\"{o}rdstrom black hole is once again that the shape dynamics solution has a wormhole character manifested by the presence of a parity horizon. The event horizon of this black hole solution is located at $r = r_* = \frac{1}{2}\sqrt{m^2-Q^2}$ where the lapse vanishes. In the limit $Q \to 0$, one recovers the uncharged spherically symmetric shape dynamic black hole and the location of the horizon reduces to $r = r_* = m/2$. Just as in the uncharged spherically symmetric black hole, the spatial metric is invariant under the parity $r \to \tilde{r} = r_*^2/r$, while the lapse changes sign under this transformation. Thus, the surface $r = r_*$ is a parity horizon. 

It is interesting to consider how the electric field $\vec{E} = E^i\partial_i$ transforms under parity. One can check that 

\begin{equation}\label{E parity}
\vec{E} = \left(1 + \frac{m}{r} + \frac{m^2-Q^2}{4r^2}\right)^{-3}\frac{Q}{r^2}\partial_r \hspace{.3cm} \to  \hspace{.3cm}
 \left(1 + \frac{m}{\tilde{r}} + \frac{m^2-Q^2}{4\tilde{r}^2}\right)^{-3}\frac{Q}{\tilde{r}^2}\partial_{\tilde{r}} = -\vec{E}
\end{equation}

\noindent so the electric field transforms like a vector under parity and transforms trivially under time reversal $t \to -t$. The lapse changes sign under time-reversal, which can be seen by noting that N can be defined through the linking theory as $N = -t^{\mu}n_{\mu}$ where $n_{\mu}$ is the unit normal\footnote{In the present setting, where the lapse can naturally take on both positive and negative values, the normal vector is taken to be fixed under time reversal.} to the maximal space-like hypersurface of constant $t$. Finally, the electric field changes sign under charge conjugation $Q \to -Q$ while the spatial metric and lapse are invariant under this transformation. Putting all this together, it is clear that the charged spherically symmetric  shape dynamic black hole solution derived above is CPT invariant.

It is worth noting that while the uncharged spherically symmetric shape dynamic black hole is fully (C)PT invariant, the rotating solution is not. It is easy to see that while the parity preserves the spatial metric and changes the sign of the lapse, the shift vector transforms like a pseudo-vector under parity (i.e. it does \emph{not} transform) while it changes sign under time-reversal. Thus, the combination of parity and time-reversal has the effect of changing the sign of the angular momentum. 

However, one can still recover PT invariance of the rotating shape dynamic black hole as an asymptotic symmetry which is approximate in the bulk but becomes exact on the horizon. This can be done by noting that the spacetime coordinate transformation $\tilde{t} =  t - \omega_0^{-1}\phi$ where $\omega_0$ is the angular velocity of the horizon measured at spatial infinity, preserves the Weyl constraint and is thus a residual gauge symmetry inherited from the linking theory. The effect of this transformation is that the new time coordinate $\tilde{t}$ is defined so that its associated coordinate derivative $\partial_{\tilde{t}} = \partial_t + \omega_0\partial_{\phi}$ is the stationarity Killing vector defining the bifurcate Killing horizon of the reconstructed spacetime. In terms of this adapted time coordinate, \eqref{Kerr PS} becomes:

\begin{equation}\label{new Kerr}
ds^2 = -\lambda^{-1}\left[d\tilde{t} + \left(\omega_0^{-1}- \omega^{-1}\right) d\phi\right]^2 + \lambda \left[m^2e^{2\gamma}(d\mu^2 + d\theta^2) + s^2d\phi^2\right].
\end{equation}

It is easy to see that in these coordinates the shift vector vanishes on the horizon, and the time reversal $\tilde{t} \to -\tilde{t}$ combined with the parity $\mu \to -\mu$ is an asymptotic symmetry of the horizon. In the static cases, there is no difference between t and $\tilde{t}$ since the angular velocity of the horizon is zero. This suggests that in defining time-reversal for stationary shape dynamic black holes, one should take $\tilde{t} \to -\tilde{t}$ rather than $t \to -t$. Based upon these considerations, one would expect that a Kerr-Newman-like shape dynamic black hole would possess an asymptotic CPT invariance that becomes exact on the horizon, but this is left for future work. 

While all of the black hole solutions I have considered are electrovac solutions, there is reason to think that parity horizons might emerge in the near horizon limit of black hole solutions of the shape dynamics equations of that contain matter in the exterior region as well. Medved, Martin and Visser have studied static and spherically symmetric but otherwise generic ``dirty" black holes \cite{MMV} in the canonical formalism\footnote{By using a time function whose associated coordinate derivative coincides with the static timelike killing vector field, the momentum conjugate to the spatial metric naturally vanishes ensuring the Weyl constraint is trivially satisfied.} by expanding in powers of the radial proper distance from the horizon. Their results show that the lapse and spatial metric define a parity horizon at least up to cubic order for lapse and up to quadratic order for the spatial metric.  It is noteworthy that this is a local construction and does not make any assumptions, such as asymptotic flatness, about boundary conditions at infinity.  

Next, I will show that Rindler space can be viewed as a solution of shape dynamics and that the Rindler horizon is also a parity horizon. 

\section{Rindler Space as a Solution of Shape Dynamics}\label{rindler}
\subsection{The Rindler Chart}
The Rindler chart represents a congruence of uniformly accelerating observers over a portion of the Minkowski spacetime. It is of particular interest because despite the fact that it is related to ordinary Cartesian coordinates in flat spacetime by a spacetime coordinate transformation, the Rindler chart contains an observer-dependent horizon which has a non-zero surface gravity, and hence a non-zero temperature, much like the event horizon of a black hole. \footnote{This is a physical effect: If one considers a Klein-Gordon field propagating in Minkowski space, one can identify the vacuum state with respect to inertial observers, but this state is not a vacuum with respect to accelerated observers---rather accelerated observers see a thermal bath of excitations of the Klein Gordon field with temperature equal to $\kappa/2\pi$ where $\kappa$ is the surface gravity of the Rindler horizon \cite{Lambert}. This is entirely analogous to what happens when one considers a scalar field in a black hole background--- initial asymptotic vacuum states are not final asymptotic vacuum states, and the result is that the black hole creates particles in the form of Hawking radiation \cite{Hawking Radiation}. The fact that accelerated observers in flat spacetime observe a thermal bath is known as the ``Unruh effect."}

In order to construct the Rindler chart, one can begin with Cartesian coordinates over Minkowski spacetime. The line element is simply:

\begin{equation}\label{Minkowski}
ds^2 = -dT^2 + dX^2 + dY^2 + dZ^2.
\end{equation}

\noindent If one then introduces the coordinate transformation

\begin{eqnarray*}
t &=& \frac{1}{\kappa}\tanh^{-1}\left(\frac{T}{X}\right) \label{t-def} \\
x &=& \sqrt{X^2-T^2} \label{x-def} \\
y &=& Y \\
z &=& Z \\
\end{eqnarray*}

\noindent one obtains the Rindler chart with the line element

\begin{equation}\label{Rindler-first}
ds^2  = -\kappa^2x^2dt^2 + dx^2 + dy^2 + dz^2.
\end{equation}

The congruence of uniformly accelerating observers see a horizon located at $x = 0$, which can be seen by noting the time-time component of the metric goes to zero there. It is worth noting that the coordinate transformation \eqref{x-def} defines $x$ only for $x > 0$. For this reason, the chart is often called the ``right Rindler wedge," and the left wedge must be defined separately with $x$ defined to have the opposite sign. Since this solution was obtained by performing a coordinate transformation that mixes space and time, it is not obvious that it is physically equivalent to the Minkowski spacetime from the point of view of shape dynamics; Spacetime diffeomorphisms are not a gauge symmetry of shape dynamics, only spatial diffeomorphisms and spatial Weyl transformations are. Note however, that this coordinate transformation preserves the Weyl constraint, so Rindler space should indeed be a solution of shape dynamics. Next, I will show how Rindler space can be derived as a solution of shape dynamics from first principles.

\subsection{Rindler Space in Shape Dynamics}
As mentioned earlier, shape dynamics trades the refoliation invariance generated by the Hamiltonian constraint of general relativity for spatial Weyl invariance generated by a new Weyl constraint. One consequence of this symmetry trading is that the lapse function is not a Lagrange multiplier in shape dynamics, but must instead satisfy  the lapse-fixing equation \eqref{LFE1}, which is reproduced below for convenience. 

\begin{equation*}
e^{-4\phi}(\nabla^2 N + 2q^{ab}\phi_{,a}N_{,b}) -
e^{-12\phi}G_{abcd}\frac{\pi^{ab}\pi^{cd}}{|q|}N = 0.
\end{equation*}
Now choose the flat initial data $q_{ij} =
\delta_{ij}$, $\pi^{ij} = 0$, and choose the gauge $\phi = 0$, so that the spatial slices in the reconstructed spacetime are flat, as opposed to conformally flat as they are in the spherically symmetric black hole cases. If the shift vector is chosen to be zero, then the reconstructed line element associated with this solution of shape dynamics will have the form:

\begin{equation}\label{static}
ds^2 = -N^2dt^2 + q_{ij}dx^idx^j.
\end{equation}

In this simple case, the lapse-fixing equation \eqref{LFE1} reduces to Laplace's equation, $\nabla^2N
= 0$. The general solution of the lapse-fixing equation for this initial data is now trivially given by the harmonic functions. In spherical coordinates, this yields:

\begin{equation}\label{harmonics}
N(r,\theta,\phi) = \sum\limits_{\ell=0}^{\infty}\sum\limits_{m=-\ell}^{\ell}\left(A_l\hspace{.05cm}r^{\ell} + B_{\ell}\hspace{.05cm}r^{-(\ell+1)} \right)Y_{\ell}^m(\theta,\phi)
\end{equation}

\noindent where $Y_{\ell}^m(\theta,\phi)$ are spherical harmonics. First, consider asymptotically flat boundary
conditions $N(\infty) = 1$ and $\partial_rN \sim \mathcal{O}(r^{-2})$, where the latter condition essentially imposes that the total energy of the solution is zero. Obviously, the only way that the first boundary condition can be satisfied is if $\ell=0$, so the solution becomes

\begin{equation}\label{harmonics plus BCs}
N(r,\theta,\phi) = \left(A_0 + \frac{B_0}{r} \right)Y_0^0(\theta,\phi).
\end{equation}

But $Y_0^0(\theta,\phi)$ is just a constant, and the second boundary condition implies that $B_0 = 0.$ Putting this together, we have $N=const.$, and since $N(\infty)=1$, this means $N=1$ everywhere. Switching back to Cartesian coordinates, one obtains the Minkowski line element:
\begin{equation}\label{minkowski}
ds^2 = -N^2dt^2 + q_{ij}dx^idx^j = -dt^2 + dx^2 + dy^2 + dz^2. \\
\end{equation} 

On the other hand, one can consider Cartesian spatial coordinates, and demand that $N(x=0) = 0$, $\frac{dN}{dx}\big|_{x=0}
= \kappa$, where $\kappa$ is a constant. It is interesting to note that this problem can be mapped onto the
problem of finding the electric potential (for $x>0$) due to an
infinite plane of surface charge at $x = 0$, so one can more or less guess the solution $N = Ex$, where is $E$ is some constant playing the role of the electric field. It is a simple exercise to show that the constant playing the role of the electric field is exactly $\kappa$, and this is carried out in the appendix for the sake of completeness.

The reconstructed line element has the form of the
Minkowski spacetime written in Rindler coordinates:

\begin{equation}\label{Rindler}
ds^2 = -N^2dt^2 + q_{ij}dx^idx^j = -\kappa^2x^2dt^2 + dx^2 + dy^2 + dz^2.
\end{equation}

It is not terribly surprising that non-inertial
observers can be obtained by changing the boundary conditions on the lapse; Eulerian
observers\footnote{Eulerian observers are congruences of timelike curves whose tangent vectors are orthogonal to the spatial hypersurfaces of a (reconstructed) foliation of spacetime. These are natural observers to choose from the point of view of shape dynamics as they are ``inertial" with respect to the spatial geometry (hypersurface orthogonal), although they are generally non-inertial in that their proper acceleration is non-zero in the reconstructed spacetime.} are generally non-inertial, with their proper acceleration
given by $a_i =\nabla_i (\ln N)$. What is more surprising is that the
spacetime reconstructions obtained from these different boundary
conditions on the lapse are related by a spacetime
diffeomorphism. The fact that this is a \emph{residual} gauge transformation from the point of view of shape dynamics may shed some light on the fact that the accelerated Eulerian observers associated with the Rindler chart observe physically different effects than their non-accelerated counterparts in the usual Cartesian Minkowski chart--- they observe a thermal bath of Unruh radiation.  Residual spacetime coordinate invariance in shape dynamics was partially explored in \cite{Lorentz}, but it is still not completely understood what role these transformations play in the theory. From the point of view of shape dynamics, these symmetries, which as in the present case may be disconnected from the identity, seem somewhat less fundamental than the gauge transformations generated  by the first class constraints which make no reference to any properties of particular solutions, and are pure-gauge even off-shell. 

Finally, it is worth noting that while at a glance the line elements seem to agree, the Rindler chart is a complete solution from the point of view of shape dynamics. That is, the coordinate $x$ is well defined for all real values and therefore covers both the right and left wedges. As a solution of shape dynamics, the Rindler chart is therefore globally different from the corresponding solution of general relativity where it is defined only for the right and left wedges separately, and globally also includes future and past wedges that are not present in the shape dynamics solution. These global differences are entirely analogous to those found when comparing the black hole solutions of shape dynamics and general relativity, but in the case of the Rindler chart, there is no event horizon. It can be easily seen that the spatial metric is invariant under the parity $x \to -x$, while the lapse changes sign under this parity, just as in the black hole solutions. Thus, the observer-dependent Rindler horizon is a parity horizon from the point of view of shape dynamics. This suggests that global disagreement with shape dynamics may arise whenever a stationary horizon of \emph{any} kind is present in a solution general relativity, and that the corresponding solution of shape dynamics will contain a parity horizon.

In the following section, I will present yet another solution of shape dynamics that possesses a parity horizon and which presents physical differences from the corresponding solution of general relativity in that the general relativistic solution contains closed timelike curves while the shape dynamics solution does not. This can be seen as an advantage for shape dynamics, as closed timelike curves violate causality and solutions of general relativity containing closed timelike curves are generally explained away on a case by case basis as being ``unphysical" for various reasons (e.g. they are quantum-mechanically unstable or they require non-localized matter distributions). In shape dynamics, no such ad hoc apologies are necessary---solutions of shape dynamics simply seem not to contain these pathologies.

\section{The Bonner Spacetime as a Solution of Shape Dynamics}\label{bonner}

\subsection{The Van Stockum Bonner spacetimes}

The Van Stockum-Bonner spacetimes are a class of asymptotically flat stationary, axisymmetric solutions of Einstein's equations sourced by rigidly rotating dust. The solutions are (with the exception of the lowest order case, which is invariant under translations parallel to the $z$-axis, and is \emph{not} asymptotically flat) ultra-relativistic in the sense that they possess no Newtonian limit. Indeed, it was pointed out in \cite{Bonner} that any density gradient in the $z$-direction would produce gravitational forces parallel to the $z$-axis that would violate the assumption of rigid rotation. In general relativity, solutions with rigidly rotating dust are possible provided the angular velocity of the dust is sufficiently rapid.

Let us briefly review the properties of the Van Stockum-Bonner spacetimes. Our discussion closely follows that of \cite{Van Stockum} which the interested reader may consult for further details.  In cylindrical coordinates, the Bonner-Van Stockum spacetimes are described by the line element

\begin{equation}\label{Van Stockum}
ds^2 = -dt^2 + 2K(\rho,z)d\phi\,dt + \left(\rho^2-K^2(\rho,z)\right) d\phi^2 + e^{2\Psi(\rho,z)}\left(d\rho^2 + dz^2 \right).
\end{equation} 

\noindent The Einstein field equations imply that

\begin{eqnarray}
\Psi_{,\rho} &=& \frac{K_{,z}^2 - K_{,\rho}^2}{4\rho} \label{EFE1} \\
\Psi_{,z} &=& -\frac{K_{,\rho}K_{,z}}{2\rho} \label{EFE2}
\end{eqnarray} 

\noindent Differentiating \eqref{EFE1} and \eqref{EFE2} with respect to $z$ and $\rho$ respectively, and noting that the partial derivatives commute then requires that $K(\rho,z)$ satisfy the linear, elliptic partial differential equation:

\begin{equation}\label{K-PDE}
K_{,\rho\rho} - \frac{1}{\rho}K_{,\rho} + K_{,zz} = 0.
\end{equation}

Given a solution of \eqref{K-PDE}, one can solve the system \eqref{EFE1}, \eqref{EFE2}  for $\Psi$ which completely determines the line element \eqref{Van Stockum}. On the other hand, it was shown in \cite{Kerr} that the general line element (up to coordinate transformations) representing a stationary, axisymmetric solution of the Einstein field equations 

\begin{equation}\label{Axi}
ds^2 = -(N^2 - \Omega\Phi\xi^2)dt^2 + \Omega[(dx^1)^2 + (dx^2)^2 + \Phi d \phi^2] + 2\Omega\Phi\xi d \phi\,dt
\end{equation}

\noindent induces a maximal slicing by hypersurfaces of constant $t$. Putting $x^1 = \rho$, $x^2 = z$, \eqref{Axi} can be put in the form of \eqref{Van Stockum} by making an appropriate choice of the metric functions $N$, $\Omega$, $\Phi$, and $\xi$:

\begin{equation}\label{metric functions}
N = \left(1 - K^2/\rho^2\right)^{-1/2}, \hspace{.5cm}
\Omega = e^{2\Psi}, \hspace{.5cm}
\Phi = e^{-2\Psi}\left(\rho^2 - K^2\right), \hspace{.5cm}
\xi = \frac{K}{\rho^2-K^2}.
\end{equation}

 This implies that \eqref{Van Stockum} induces a maximal slicing by hypersurfaces of constant $t$, which is precisely the criterion needed to map a solution of general relativity onto a solution of shape dynamics, at least locally.  Every solution of \eqref{K-PDE} yields a solution of \eqref{Van Stockum} that possesses a surface defined by the equation $\rho^2 = K^2$, and it can be readily seen that such a surface has two important properties. 

First, on such a surface $g_{\phi\phi}=0$, and it can be easily checked for any solution that this divides the spacetime into two regions: an ``exterior" region defined by $\rho^2>K^2$ and an ``interior region" defined by $\rho^2<K^2$. In the interior region, $g_{\phi\phi}<0$, and the coordinate basis vector $\partial_{\phi}$ is therefore \emph{timelike} in this region. Since the coordinate $\phi$ is periodic with period $2\pi$, the integral curves of $\partial_{\phi}$ are closed. Therefore, the interior region is filled with closed timelike curves. Any surface which divides a spacetime into a causal region without closed timelike curves, and an acausal region with closed timelike curves is a special kind of Cauchy horizon called a chronological horizon \cite{Thorne}.

The second important feature of this chronological horizon, is that the lapse function, $N$ goes to infinity there, as can be easily from the first of equations \eqref{metric functions}. Recall from \ref{Background} that the determinant of the spacetime metric can be written $\sqrt{-g} = N\sqrt{q}$, and if this quantity is finite as it must be for a non-singular region of spacetime, then if the lapse function $N$ diverges on some surface, the determinant of the spatial metric must vanish. This is indeed the case, since $g_{\phi\phi}=q_{\phi\phi}=0$ and there are no off-diagonal terms in the spatial metric. This is a signal that the ``spatial" metric changes signature across this surface and is no longer truly spatial in the interior region. This is physically unacceptable from the point of view of shape dynamics, since shape dynamics is a theory of the time evolution of conformal equivalence classes of \emph{Riemannian} three-manifolds. Nevertheless, the fact that the line element \eqref{Van Stockum} induces a maximal slicing by hypersurfaces of constant $t$ in the exterior  region, means that this is a solution of shape dynamics in the exterior region. In what follows, I will consider a spatial coordinate transformation that is a diffeomorphism only in the exterior region, but which takes the line element \eqref{Van Stockum} into a form that makes the spatial metric non-degenerate across the chronological horizon, yielding a complete solution of shape dynamics that covers the whole spatial manifold. 

\subsection{The Bonner Solution for Shape Dynamics}

Equation \eqref{K-PDE} is linear and invariant under translations in $z$-direction. Consequently, any linear combination of known solutions and their $z$-derivatives of any order is again a solution. A simple class of solutions can be obtained by expanding $K(\rho,z)$ in a superposition of modes. The external multipolar solutions are given by 

\begin{equation}\label{modes}
  K(\rho,z)=\begin{cases}
   z\left(\rho^2 + z^2\right)^{-\frac{n+1}{2}}\hspace{.22mm}_2F_1\left(\frac{n+1}{2},-\frac{n}{2};\frac{3}{2};\frac{z^2}{\rho^2+z^2}\right), & \text{$n=0,2,4,6...$}\\
\hspace{2.4mm}  \left(\rho^2 + z^2\right)^{-n/2}\hspace{1mm}_2F_1\left(-\frac{n+1}{2},\frac{n}{2};\frac{1}{2};\frac{z^2}{\rho^2+z^2}\right)  , & \text{$n=1,3,5,7...$}
  \end{cases}
\end{equation}

\noindent where $\hspace{.22mm}_2F_1(a,b;c;\eta)$ is the \emph{hypergeometric function} defined for $|\eta|<1$ by the infinite series

\begin{equation}\label{hypergeo}
\hspace{.22mm}_2F_1(a,b;c;\eta) = \sum\limits_{m=0}^{\infty}\frac{(a)_m(b)_n}{(c)_m}\frac{\eta^m}{m!}
\end{equation}

 \noindent and $(q)_m$ is the Pochhammer symbol:

\begin{equation}\label{weird thing}
(q)_m = \begin{cases}
1, &\text{$m=0$} \\
q(q+1)...(q+m-1), &\text{$m>0$}.
\end{cases}
\end{equation}

\noindent When either $a$ or $b$ is a non-positive integer, as it is in the expressions for the multipolar solutions for $K(\rho,z)$, the hypergeometric function becomes a polynomial of finite order:

\begin{equation}\label{hypergeo-finite}
\hspace{.22mm}_2F_1(-n,b;c;\eta) = \sum\limits_{m=0}^{n}(-1)^m\colvec{n}{m}\frac{(b)_m}{(c)_m}\eta^m.
\end{equation}  

\noindent For simplicity, I will consider a particular solution to \eqref{K-PDE} and its associated line element \eqref{Van Stockum} corresponding to the $n=1$ mode of \eqref{modes}, for which $K(\rho,z) = \-2h\frac{\rho^2}{\left(\rho^2 + z^2\right)^{3/2}}$ and $\Psi(\rho,z) = \frac{h^2}{4}\frac{\rho^2\left(\rho^2-8z^2\right)}{\left(\rho^2 + z^2\right)^4}$ where $h$ is a positive constant with dimensions of area parametrizing the location of the chronological horizon. Details about this solution can be found in \cite{Bonner}. 

It is more convenient to work in spherical coordinates, where $K(r,\theta) = \frac{2h}{r}\sin^2\theta$, and \\ $\Psi(r,\theta) = \frac{h^2}{4}r^{-4}\sin^2\theta\left(\sin^2\theta - 8\cos^2\theta\right)$. After changing to spherical coordinates, the line element \eqref{Van Stockum} becomes

\begin{equation}\label{VanStockSphere}
ds^2 = -dt^2 +2Kd\phi dt + e^{2\Psi}\left(dr^2 + r^2d\theta^2\right) +\left(r^2\sin^2\theta-K^2\right)d\phi^2.
\end{equation}

This solution is known as the Bonner spacetime, and in spherical coordinates, the equation defining the chronological horizon is given by

\begin{equation}\label{horizon}
 r^2\sin^2\theta - \frac{4h^2}{r^2}\sin^4\theta = g_{\phi\phi} = 0 \hspace{5mm}
\text{or} \hspace{5mm}
r^2 = 2h\sin\theta.
\end{equation}

\begin{figure}[h]
\centering
\includegraphics[width = 7.5cm, height = 5.5cm]{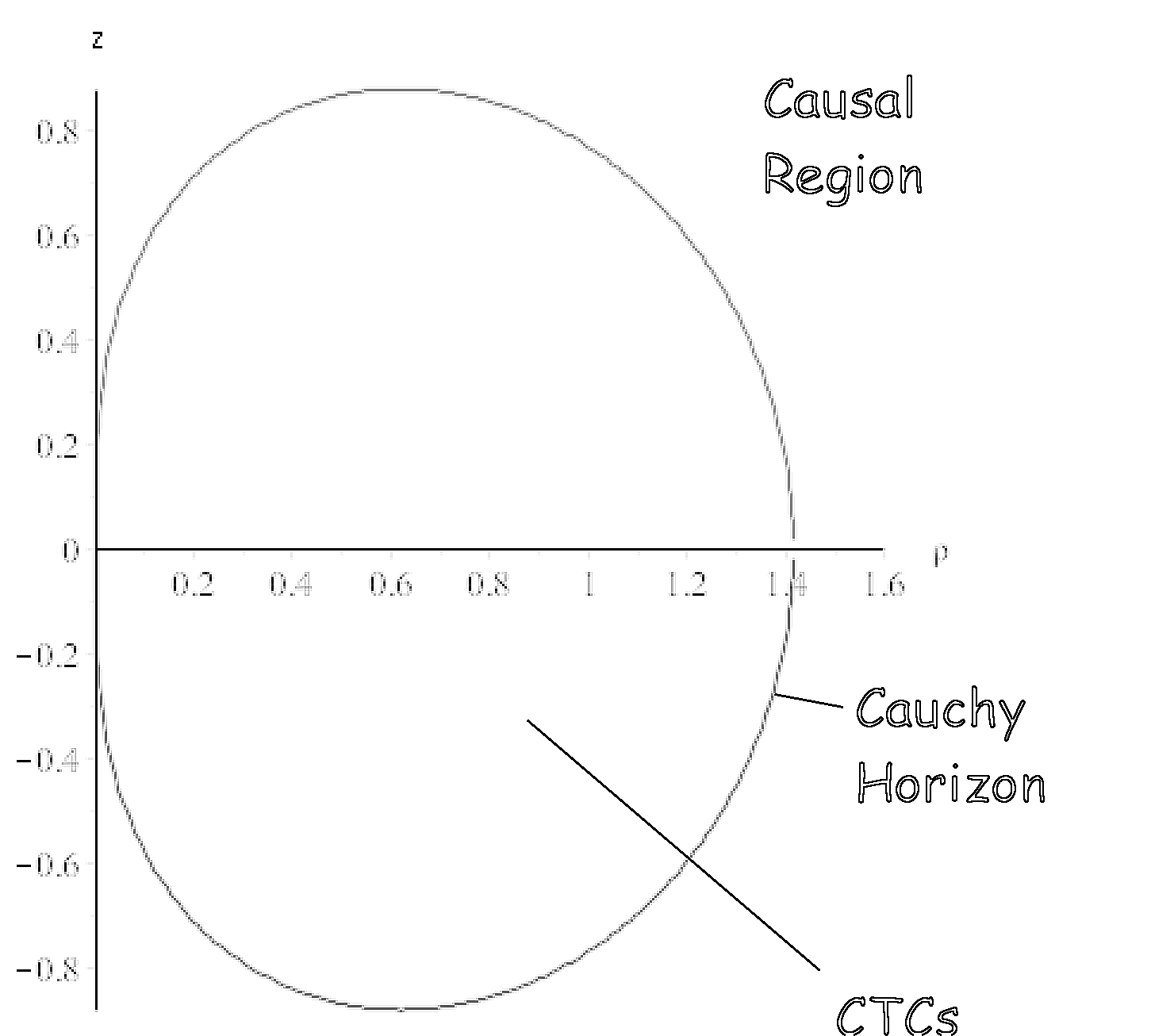} \\
\caption{\footnotesize A plot of the chronological horizon horizon in the Bonner spacetime for $h = 1$.}
\label{rosy plane}
\end{figure}

The chronological horizon in the Bonner spacetime has roughly the shape of a degenerate torus, with its inner circumference shrunken to a point. \ref{rosy plane} depicts the chronological horizon in the $\rho\hspace{.3mm}$-$z$ half plane.

Equation \eqref{horizon} depends on both $r$ and $\theta$. To find coordinates that make the spatial metric smooth across the chronological horizon, it will be convenient to work in coordinates that are adapted to this surface, so that \eqref{horizon} takes the form $x = constant$ for some new spatial coordinate $x$. This can be accomplished by making the coordinate transformation

\begin{eqnarray}\label{to u and v}
x &=& \sqrt{r^2-2h\sin\theta} \\
y &=& \sqrt{r^2+2h\sin\theta}
\end{eqnarray}

\noindent In the $(x,y)$ system, the chronological horizon is located at $x=0$, and the spatial part of the line element becomes

\begin{eqnarray}\label{xy-element}
d\ell^2 &=& q_{ij}dx^idx^j \nonumber \\ &=& e^{2\Psi}\left[\frac{\frac{1}{2}\left(xdx+ydy\right)^2}{x^2+y^2} + \frac{1}{8h^2}\left(x^2+y^2\right)\frac{\left(y^2-x^2\right)^2}{\left(y^2-x^2\right)^2-16h^2}\left(ydy-xdx\right)^2\right]+\frac{x^2y^2}{32h^2}\frac{\left(y^2-x^2\right)^2}{x^+y^2}d\phi^2. \nonumber \\
\,
\end{eqnarray}

\noindent In the near-horizon limit $x^2<<y^2$, \eqref{xy-element} reduces to

\begin{equation}\label{near-horizon}
d\ell^2 \approx e^{2\Psi}\left[Q_+(y)\left(x^2dx^2 + y^2dy^2\right) + 2Q_-(y)xy\,dx\,dy\right] + \frac{x^2y^4}{32h^2}d\phi^2
\end{equation}

\noindent where I have defined 

\begin{equation}\label{Q-def}
Q_{\pm}(y) = \frac{1}{2y^2} \pm \frac{1}{8h^2}\frac{y^4}{y^4-16h^2}.
\end{equation}

\noindent From equation \eqref{near-horizon} one can read off the determinant of the spatial metric in the near horizon limit:

\begin{equation}
|q| = e^{4\Psi}\left(Q_+^2-Q_-^2\right)\frac{x^4y^6}{32h^2}.
\end{equation}

Clearly, as $x \to 0$, $|q| \to 0$ as well, indicating that the spatial metric is still degenerate across the chronological horizon in the $(x,y)$ system. This can be remedied by making one final spatial coordinate transformation that is a diffeomorphism only for $x>0$. Since the determinant of the metric transforms as $|q| \to |\overbar{q}|=|J|^2|q|$ under a coordinate transformation $dx^b \to d\overbar{x}^a = J^a_{\,\,\,b} \,dx^b$ with Jacobian $J^a_{\,\,\,b} = \frac{\partial \overbar{x}^a}{\partial x^b}$, one can make the near-horizon spatial geometry non-degenerate by transforming the $x$ coordinate so that the $x$-dependence of the near-horizon determinant of the spatial metric is canceled. This can be done by choosing a new coordinate $u$ so that $\frac{du}{dx} \propto x^2$. Making an analogous transformation of the $y$ coordinate to preserve the symmetry of the metric, one can choose:

\begin{eqnarray}
u &=& h^{-1}x^3 \label{x to u}, \\
v &=& h^{-1}y^3 \label{y to v}.
\end{eqnarray}

\noindent where the prefactors of $h^{-1}$ appear so that the new coordinates have dimensions of length. In the $(u,v)$ system, the spatial line element becomes

\begin{eqnarray}\label{uv-element}
d\ell^2 &=& \overbar{q}_{ij}d\overbar{x}^id\overbar{x}^j \nonumber \\ &=& e^{2\Psi}\left[\overbar{Q}_+(u,v)\left(u^{-2/3}du^2 + v^{-2/3}dv^2\right) + 2\overbar{Q}_-(u,v)u^{-1/3}v^{-1/3}du\,dv\right] + \frac{u^{2/3}v^{2/3}}{32}\frac{\left(v^{2/3}-u^{2/3}\right)^2}{u^{2/3}+v^{2/3}}d\phi^2 \nonumber \\
\,
\end{eqnarray}

\noindent where I have defined 

\begin{equation}\label{Q-bar}
\overbar{Q}_{\pm}(u,v) = \frac{1}{18}\left[\frac{\alpha^2}{u^{2/3}+v^{2/3}} \pm \left(u^{2/3}+v^{2/3}\right)\frac{\left(v^{2/3}-u^{2/3}\right)^2}{\left(v^{2/3}-u^{2/3}\right)^2-16\alpha^2}\right], \hspace{3mm} \alpha = h^{1/3}. 
\end{equation}

\noindent In the near horizon limit $u^{2/3}<<v^{2/3}$, one finds that 

\begin{equation}\label{Q-bar limit}
\overbar{Q}_{\pm}(u,v) \approx \overbar{Q}_{\pm}(v) = \frac{1}{18}\left[\frac{\alpha^2}{v^{2/3}} \pm v^{2/3}\frac{v^{4/3}}{v^{4/3}-16\alpha^2}\right]
\end{equation}

\noindent and the near-horizon limit of the spatial determinant becomes 

\begin{equation}\label{near-horizon uv det} 
|\overbar{q}| \hspace{1mm} \approx \frac{e^{4\Psi}}{32}\left(\overbar{Q}_+(v)^2-\overbar{Q}_-(v)^2\right)\frac{\left(v^{2/3}-u^{2/3}\right)^2}{u^{2/3}+v^{2/3}}
\end{equation}

\noindent which is finite and non-zero on the chronological horizon $u=0$. 

Clearly, the transformation \eqref{x to u} is a diffeomorphism only for $u>0$, i.e. outside the chronological horizon, as its inverse transformation $x = (hu)^{1/3}$ is not differentiable at $u = 0$. Nevertheless, the spatial metric is non-degenerate across the chronological horizon, and the coordinate $u$ can be extended to all real values. Noting that the line element \eqref{uv-element} is invariant under $u \to -u$, it can easily be seen that the spatial geometry possesses a reflection symmetry about the chronological horizon. Moreover, in the $(u,v)$ system, the lapse becomes

\begin{equation}
N(u,v) = \frac{u^{2/3} + v^{2/3}}{(uv)^{1/3}}
\end{equation}

\noindent which changes sign under the parity $u \to -u$ . Just as in the other solutions of shape dynamics with horizons that have been considered, the ``interior" region may be interpreted as a time-reversed copy of the exterior region--- the chronological horizon becomes a parity horizon. This makes it obvious that no portion of the reconstructed spacetime possesses closed timelike curves. Finally, it is worth emphasizing that since the spatial geometry is non-degenerate for all values of $u$, and since the line element \eqref{uv-element} is globally related to \eqref{Axi} by a spatial diffeomorphism (except on the horizon, which needs to be treated with some care), this line element corresponds to a \emph{complete} solution of shape dynamics, covering the entire spatial manifold except the horizon, $u = 0$ which will be discussed in the following section.

\section{Conformal Singularity of the Chronological Horizon}

In the previous section, it was shown that in the $(u,v)$ system the spatial metric \eqref{uv-element} becomes non-degenerate and the chronological horizon becomes a parity horizon. The canonical momentum in these coordinates can be found by considering Hamilton's equation for $\dot{q}_{ij}$\footnote{The bar over the metric has been omitted for notational convenience.}:

\begin{equation}\label{H1}
\dot{q}_{ij} = 4\rho(x) q_{ij} + 2e^{-6\phi(x)}\frac{N}{\sqrt{q}}\left(\pi_{ij} - \frac{1}{2}\pi q_{ij}\right) + \mathcal{L}_{\xi}q_{ij}.
\end{equation}

\noindent where $\mathcal{L}_{\xi}q_{ij}$ denotes the Lie derivative of the spatial metric along the shift vector. With the gauge fixing conditions $\rho(x) = 0$, $\phi(x) = 0$, and noting that $\dot{q}_{ij} = \partial_{\phi}q_{ij} = 0$, \eqref{H1} yields the simple expression

\begin{equation}\label{pi}
\pi_{ij} = -\frac{\sqrt{q}\hspace{.3mm}q_{\phi\phi}}{N}\left(\delta^u_{(i}\delta^{\phi}_{j)}\xi_{,u} + \delta^v_{(i}\delta^{\phi}_{j)}\xi_{,v}  \right)
\end{equation}

\noindent for the canonical momentum $\pi_{ij}$. It is easily seen from \eqref{H1} that the trace of the momentum $\pi = 0$ and this can be explicitly confirmed from \eqref{pi} and \eqref{uv-element}, so the Weyl constraint is satisfied. One can also check that as $u \to 0$, $\pi_{u\phi} \sim u^{-2/3} \to \infty$. This is not surprising as the transformation to the $(u,v)$ system deliberately introduced a singularity in the $q_{uu}$ and $q_{uv}$ components of the spatial metric in order to cancel the degeneracy from the vanishing the of $q_{\phi\phi}$ component. Next, it will be shown that the singularities appearing in spatial metric and its conjugate momentum are of a physical nature--- i.e. the chronological horizon of the Bonner solution for shape dynamics is an extended physical singularity. Finally, we argue that this result can generically be extended to a broader class of solutions containing chronological parity horizons. 

Since shape dynamics is a theory of evolving conformal geometries, our analysis will focus on invariants constructed from the conformally invariant degrees of freedom of the spatial curvature. In three spatial dimensions, the rank-three Cotton tensor

\begin{equation}\label{Cotton1}
\mathscr{C}_{ijk} := \nabla_k\left(R_{ij} - \frac{R}{4}q_{ij}\right) - \nabla_j\left(R_{ik} - \frac{R}{4}q_{ik}\right)
\end{equation}

 \noindent contains all of the local information concerning the conformal geometry of space with the Cotton tensor vanishing if and only if the spatial geometry is conformally flat. The algebraic complexity of the line element \eqref{uv-element} makes computation of the Cotton tensor difficult, but one can analyze the behavior of the spatial conformal geometry of the exterior by considering the Cotton tensor in spherical coordinates for $h=1$. When this is done, one can compute the square of the Cotton tensor:

\begin{equation}\label{C-squared spherical}
\mathscr{C}^{ijk}\mathscr{C}_{ijk} = 4r^{-22}\exp\left(\frac{3}{2}r^{-4}\sin^2\theta\left(3\cos\theta+1\right) \left(3\cos\theta-1\right)\right)\frac{A(r,\cos\theta)}{B(r,\cos\theta)}
\end{equation}

\noindent Where $A(r,\cos\theta)$ is a polynomial of degree 32 in $r$ and $\cos\theta$ with integer coefficients and $B(r,\cos\theta)$ is a polynomial of degree 24 in $r$ and $\cos\theta$ with integer coefficients. It can be shown using numerical methods that $A(r,\cos\theta)$ is non-zero on the horizon, while $B(r,\cos\theta)$ goes to zero there. Plots of $B(r,\theta)$ are shown below. 

\begin{figure}[h]
\centering
\includegraphics[width = 7.5cm, height = 5.5cm]{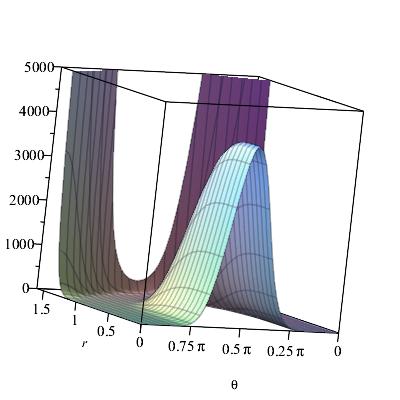} \\
\caption{\footnotesize A plot of $B(r,\theta) < 5000$ for $h = 1$ with $0 < r < 1.6$, $0 < \theta < \pi$.}
\label{3d1}
\end{figure}

\begin{figure}[h]
\centering
\includegraphics[width = 7.5cm, height = 5.5cm]{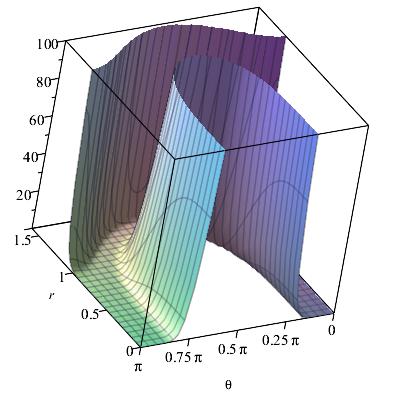} \\
\caption{\footnotesize A plot of $B(r,\theta) < 100$ for $h = 1$ with $0 < r < 1.6$, $0 < \theta < \pi.$}
\label{3d2}
\end{figure}

\ref{3d1} shows the general behavior of $B(r,\theta)$ near the horizon, while the range in \ref{3d2} is restricted to $[0,100]$ in order to display the shape of the horizon in the $r-\theta$ plane. One can see from figures 2 and 3 that $B(r,\theta)$ goes smoothly to zero along a curve in the $r-\theta$ plane, and it can be shown that this curve exactly coincides with the chronological horizon $r^2 = 2h\sin\theta.$ For values of $r$ and $\theta$ that lie in the interior of the horizon, the plots have no physical significance, as the $(r,\theta)$ system yields an ill-defined spatial conformally geometry for these values. Nevertheless, we can gain insight from the behavior of the invariant $\mathscr{C}^{ijk}\mathscr{C}_{ijk}$ outside and on the horizon. Since this invariant diverges smoothly as one approaches the horizon, and since it is an invariant of the spatial conformal geometry, there exists no spatial diffeomorphism of the exterior that can remove this divergence. Thus, we are forced to conclude that the cost of removing the degeneracy of the spatial metric in order to find a complete solution of shape dynamics is that the parity horizon of the shape dynamics solution must be regarded as an extended physical singularity. It is noteworthy that this is the first known solution of shape dynamics that possesses a ``shape singularity"---i.e. a singularity in the spatial conformal geometry.  

There is good reason to think that the feature we have just observed, namely that the chronological horizon becomes an extended physical singularity in shape dynamics, holds quite a bit more generally than in the specific case we have considered above. If we consider the spatial metric associated with the line element \eqref{Axi}, we can compute the general form of the invariant $\mathscr{C}^{ijk}\mathscr{C}_{ijk}$ in cylindrical coordinates:

\begin{equation}\label{C-squared}
\mathscr{C}^{ijk}\mathscr{C}_{ijk} = \frac{1}{ 4\psi^
{6} \Omega^{3}} 
\Bigg[ \psi^{2}{
\frac {\partial ^{3}\psi}{\partial {z}^{3}}} +
\psi^{2}{\frac {\partial ^{3}\psi}{
\partial {\rho}^{2}\partial z}}-2\,  \psi{\frac {\partial\psi }{\partial z}}{
\frac {\partial ^{2}\psi}{\partial {z}^{2}}}
 -\hspace{.5mm}2\, \psi {
\frac {\partial \psi}{\partial \rho}}{\frac {\partial ^{2}\psi}{\partial \rho
\partial z}}  + {\frac 
{\partial\psi }{\partial z}}\left( {
\frac {\partial\psi }{\partial \rho}}\right) ^{2}+
 \left( {\frac {\partial\psi }{\partial z}}
 \right) ^{3} \Bigg] ^{2} +\,\, (\rho \leftrightarrow z) 
\end{equation}

Since the chronological horizon for any such solution is defined by $\psi(\rho,z)=0$, we see that the invariant generically diverges for any stationary, axisymmetric solution of shape dynamics with a chronological horizon so long as the numerator of the invariant does not conspire to cancel this divergence.

\section{Discussion}

The various known and novel stationary, axisymmetric solutions of asymptotically flat shape dynamics presented in this work demonstrate the pervasiveness of parity horizons in this context. The fact that black hole solutions of shape dynamics are physically different from those of general relativity opens up new possibilities for black hole physics and thermodynamics if shape dynamics is the correct description of the true degrees of freedom of the classical gravitational field. Some very preliminary results on the thermodynamics of shape dynamic black holes was discussed in \cite{Vasu}. A particularly striking feature of these solutions is that they contain no central physical singularity, a fact that strongly suggests that shape dynamic black holes do not suffer from an information loss paradox as general relativistic black holes do.  The CPT invariance of electrovac black hole solutions with a high degree of symmetry presents an interesting possible connection to the standard model of particle physics. While shape dynamic black holes have some promising features, there are still some important open questions that need to be addressed, not least of which is whether such solutions can form from collapse, a problem which is already being explored and which promises to shed light on the known eternal solutions. Collapse models are also an important step in understanding two features of shape dynamic black holes which are currently poorly understood: First, what does an observer see as she passes through the horizon of a shape dynamic black hole? And second, do shape dynamic black holes evaporate via Hawking radiation? 

The first of these questions arises as a result of the fact that these solutions do not form well-defined spacetime geometries on the horizon. As a result, the timelike component of the geodesic equation is ill-defined at the horizon, and the trajectories of observers that would ordinarily be expressed as timelike geodesics con no longer be described in this manner. This might be resolved by defining timelike geodesics connecting points in the interior and exterior at different times as piece-wise smooth curves that are timelike geodesics in the interior and exterior regions and that minimize the proper time. In this way the resulting degenerate spacetime becomes geodesically complete, although the geodesics defined in this manner have a discontinuity in their tangent vector at the horizon. While this seems like a reasonable definition, it remains troubling that the trajectories of observers cannot be determined locally as the solution of a second or differential equation that is well-defined everywhere.

The second question was partially addressed in \cite{Vasu} where it was noted that since shape dynamic black holes agree with general relativistic black holes in the exterior region it may be possible to import whole-cloth certain derivations of Hawking radiation that make no reference to the interior region. This can be done by solving the Klein-Gordon equation in the background of the exterior black hole geometry, quantizing the Klein-Gordon field, and relating the ingoing modes at past null infinity to ingoing modes at the horizon and outgoing modes at future null infinity via Bogoliubov transformation \cite{Lambert}. Still, it would be preferable to derive Hawking radiation in the context of a collapse model in order to obtain a more concrete physical picture of the dynamics of black hole formation and evaporation.  

Rindler space provides a simple example of a solution of shape dynamics that might be used to study black hole thermodynamics via the Unruh effect. Moreover, all of the black hole solutions of shape dynamics share the same global causal structure as Rindler space, so this simple example might be used as a toy model to study more complicated questions about shape dynamic black holes. 

The singular parity horizon encountered in the case of the Bonner solution clarifies the role of the assumption of global hyperbolicity in shape dynamics, and suggests a general chronology protection mechanism in shape dynamics. One can assume global hyperbolicity from the outset and then solutions with closed timelike curves are automatically excluded, but there remain solutions of general relativity which admit maximal slicing outside of the chronological Cauchy horizon. The procedure outlined in the case of the Bonner spacetime explains how these two facts can be reconciled. By making a singular coordinate transformation that makes the spatial metric invertible across the Cauchy horizon, one obtains a global solution of shape dynamics in which closed timelike curves do not form, and which contains closed null curves only on a singular sub-surface of measure zero. If one demands continuity of the phase space variables then these solutions must be discarded, since the metric and momentum are divergent there and the singularity in the square of the Cotton tensor ensures that there is no spatial diffeomorphism which can remove these divergences.

Clearly, much work remains in understanding the physical differences between shape dynamics and general relativity, but the solutions presented in this work suggest that parity horizons, or some dynamical generalization thereof my be of great utility in characterizing the possible global differences between the two theories that can arise.

\subsection*{Acknowledgments}
I am grateful to Steve Carlip for carefully reading drafts and making many helpful suggestions, and to Joseph Mitchell for many lively conversations about this work as it progressed. I would also like to thank Jack Gegenberg for helpful discussions in the early stages of this work, and Henrique Gomes, Sean Gryb, Flavio Mercati and Vasudev Shyam for their thoughtful questions and remarks.
\section*{Appendix}
The general solution to Laplace's equation in Cartesian coordinates can be written:

\begin{gather}
\nonumber \\
N(x,y,z) = \sum\limits_{n_y,n_z =0}^{\infty}\left(A_{x,n_y,n_z} e^{\frac{x}{2\pi}\sqrt{n_y^2 -n_z^2}} + B_{x,n_y,n_z} e^{-\frac{x}{2\pi}\sqrt{n_y^2 -n_z^2}}\right) \nonumber\\ \times \left(A_{y,n_y}\sin\left(\frac{n_y y}{2\pi}\right)+B_{y,n_y}\cos\left(\frac{n_y y}{2\pi}\right)\right) \label{rectangular harmonics}\\ \times \left( A_{z,n_z}\sin\left(\frac{n_z z}{2\pi}\right)+B_{z,n_z}\cos\left(\frac{n_z z}{2\pi}\right)\right). \nonumber
\end{gather}

\noindent Imposing $N(0,y,z) = 0$, one obtains $A_x + B_x = 0$, so the solution \eqref{rectangular harmonics} becomes:

\begin{gather}
N(x,y,z) = \sum\limits_{n_y,n_z =0}^{\infty}2A_{x,n_y,n_z}\sinh\left(\frac{x}{2\pi}\sqrt{n_y^2 -n_z^2}\right) \nonumber \\
\times\left(A_{y,n_y}\sin\left(\frac{n_y y}{2\pi}\right)+B_{y,n_y}\cos\left(\frac{n_y y}{2\pi}\right)\right) \label{rectangular harmonics + BC1}\\
\times\left(A_{z,n_z}\sin\left(\frac{n_z z}{2\pi}\right)+B_{z,n_z}\cos\left(\frac{n_z z}{2\pi}\right)\right). \nonumber \\
\nonumber
\end{gather}

\noindent If we then impose 

\begin{equation}\label{firstDer}
\pder{N}{x}\bigg|_{x=0} = \kappa 
\end{equation}

\noindent the only allowed values of $n_y$ and $n_z$ are zero, so the A's and B's corresponding to $n_y, n_z \neq 0$ must all vanish, and one obtains

\begin{equation}\label{BC2}
\kappa = \frac{c_{n_y,n_z}}{2\pi}\sqrt{n_y^2 -n_z^2}\bigg|_{n_y=n_z=0}
\end{equation}

\noindent where $c_{n_y,n_z} = 2A_{x,n_y,n_z}B_{y,n_y}B_{z,n_z}$. Solving \eqref{BC2} for $c_{n_y,n_z}$ and inserting the result back into \eqref{rectangular harmonics + BC1} gives

\begin{equation}\label{limit1}
N(x) = \frac{2\pi\kappa}{\sqrt{n_y^2 -n_z^2}}\sinh\left(\frac{x}{2\pi}\sqrt{n_y^2 -n_z^2}\right)\bigg|_{n_y=n_z=0}
\end{equation}

\noindent Clearly, this equation must be interpreted as a limit, which can be evaluated using L'Hospital's rule for indeterminate forms:

\begin{eqnarray}\label{limit 2}
N(x) = \lim_{a \to 0}\frac{2\pi\kappa}{a}\sinh\left(\frac{ax}{2\pi}\right)
= \lim_{a \to 0}\frac{2\pi\kappa}{1}\frac{x}{2\pi}\cosh\left(\frac{ax}{2\pi}\right) = \kappa x.  
\end{eqnarray}

\end{document}